# High-magnetic-field Tunneling Spectra of ABC-Stacked Trilayer Graphene

Long-Jing Yin[1,2], Li-Juan Shi[2], Si-Yu Li[1], Yu Zhang[1], Zi-Han Guo[1], and Lin He[1,*]

[1] *Center for Advanced Quantum Studies, Department of Physics, Beijing Normal University, Beijing, 100875, China*

[2] *School of Physics and Electronics, Hunan University, Changsha, 410082, China*

*Corresponding author: helin@bnu.edu.cn

**ABC-stacked trilayer graphene (TLG) are predicted to exhibit novel many-body phenomena due to the existence of almost dispersionless flat-band structures near the charge neutrality point (CNP). Here, using high magnetic field scanning tunneling microscopy, we present Landau Level (LL) spectroscopy measurements of high-quality ABC-stacked TLG. We observe an approximately linear magnetic-field-scaling of the valley splitting and orbital splitting in the ABC-stacked TLG. Our experiment indicates that the valley splitting decreases dramatically with increasing the LL index. When the lowest LL is partially filled, we find an obvious enhancement of the orbital splitting, attributing to strong many-body effects. Moreover, we observe linear energy scaling of the inverse lifetime of quasiparticles, providing an additional evidence for the strong electron-electron interactions in the ABC-stacked TLG. These results imply that interesting broken-symmetry states and novel electron correlation effects could emerge in the ABC-stacked TLG in the presence of high magnetic fields.**

Searching for systems with low energy flat bands has attracted considerable interest in condensed matter physics and other fields due to the important role played by electron-electron interaction in the flat bands [1-3]. In such systems, electrons become superheavy in the dispersionless flat bands and their Coulomb energy can be far greater than the kinetic energy. Consequently, electronic interactions grow much stronger and various exotic correlated phenomena, including fractional quantum Hall effect, ferromagnetism, and superconductivity, are expected to develop. Several systems, such as kagome [1], Lieb [2], and checkerboard lattices [3], and magic-angle twisted graphene bilayers [4-6], have been proposed to have flat bands. Very recently, several experiments really observed novel many-body phenomena in the solid-state materials with flat bands [1,7-9]. However, all these systems need specific design of the lattice geometry. The ABC-stacked trilayer graphene (TLG) exists naturally and has nearly flat bands around the charge neutrality point (CNP) due to the distinct energy dispersion of $E \sim k^3$ (where $k$ is the wave vector) [10]. Therefore, it has drawn intense attention recently [11-20]. Experimentally, despite the low energy flat-band structures in ABC-stacked TLG has been clearly proved [21-24], the underlying electron-electron interactions remain relatively unexplored. Until very recently, a transport measurement of ABC TLG/boron nitride heterostructures observed tunable Mott insulator states at 1/4 and 1/2 fillings of the miniband in zero magnetic field [25], suggesting that novel strongly correlated phenomena could be realized in the ABC-stacked TLG when its bands are partially filled [18-20].

In this letter, we report the unambiguous observations of interesting broken-symmetry states and many-body effects in the quantum Hall regime of ABC-stacked TLG by probing the Landau level (LL) spectroscopy with high-magnetic-field scanning tunneling microscopy/spectroscopy (STM/STS) experiments. In the presence of a magnetic field, energy bands of the ABC TLG develop into more flat LLs, with each level has fourfold degeneracy (dual spin and dual valley) except the lowest one (with extra orbital-triplet). As a result, electron-electron interactions become more pronounced, giving rise to large splitting of the degenerate LL, especially for the 12-fold degenerate lowest LL of the ABC TLG. Our LL spectroscopy measurements explicitly demonstrate such level splitting with different broken degeneracies in the ABC-stacked TLG.

Our experiments are performed on electronically decoupled TLG on graphite substrate by using a high-magnetic-field (0-15 T) STM at $T = 4.2$ K (for a detailed description of the measurement technique, see supporting materials [26]). Figure 1(a) shows a STM topographic image around a step edge with one-layer height [see lower inset of Fig. 1(a)]. Our high-field STS measurements demonstrated that the left region of the step edge is a decoupled ABC-stacked TLG region and the right region of the step edge is a decoupled AB-stacked bilayer. The zero-field STS spectrum, *i.e.*, the differential conductivity (*dI/dV*) spectrum, recorded on the AB-stacked region shows a finite gap ~30 meV [Fig. 1(c)], as observed previously in AB-stacked bilayer on graphite [27-29]. The spectrum recorded on the ABC TLG region exhibits a pronounced density-of-state (DOS) peak near the Fermi energy, as shown in Fig. 1(b). The STS spectrum reflects the local DOS of electrons beneath the STM tip, therefore, the DOS peak observed in the ABC region is attributed to the flat bands at the CNP of the topmost ABC-stacked TLG [21,22]. It's worth noting that no clear energy gap is found among the flat-band peak under a ~1 meV energy resolution of the spectra. We find a striking homogeneity of energy of the flat bands (see Fig. S1 in supporting materials [26]), demonstrating the ultra-low charge fluctuations in our sample. Moreover, the peak width (~ 6 meV) of the DOS peak is much smaller than that of the flat bands measured in the magic-angle twisted bilayer graphene (~ 18 meV) under the same experimental conditions [6], implying that the band structures near the CNP are more flat, i.e., the kinetic energy of quasiparticles in the flat bands is much smaller, in our ABC TLG system.

The ABC stacking order and high-quality of the TLG sample can be unambiguously verified by LL spectroscopy measurements under various perpendicular magnetic fields $B$ [21,29]. Figure 1(d) shows the LL spectroscopy for the ABC TLG measured from 0 – 15 T. A series of well-separated LLs are resolved both in the electron and hole sides, suggesting that the TLG is effectively decoupled from the substrate and has a very high-quality [30]. The distinct Landau quantization and band structure information of the

ABC-stacked TLG then can be extracted from the measured LL spectroscopy. From the simplest tight-binding model including only nearest intralayer and interlayer coupling ($\gamma_0$ and $\gamma_1$), the LL spectrum for ABC-stacked TLG is expressed as [31,32]

$$E_n = E_C \pm \frac{2\hbar v_F^2 eB^{3/2}}{\gamma_1^2} \sqrt{n(n-1)(n-2)}, \qquad (1)$$

Where $n$ = integer is the LL index, $E_C$ is the energy of the CNP, $\pm$ denotes electrons and holes, $e$ is the unit charge, $\hbar$ is the Plank constant, and $v_F \sim 10^6$ m/s is the Fermi velocity. Obviously, there is a $B$-independent lowest LL at $E_C$ that is degenerated by $n$ = 0, 1, and 2 orbits. Besides, $E_n$ of other LLs is proportional to $[n(n-1)(n-2)B^3]^{1/2}$ [see inset of Fig. 1(e)] arises from the distinct cubic dispersion $E = \pm v_F^3 k^3 / \gamma_1^2$ for ABC-stacked TLG. The momentum $k_n$ for the $n$th LL at energy $E_n$ can be derived from the semi-classical Onsager quantization condition [33]. This condition specifies that the area $S_n$ of the $n$th LL in $k$-space has to quantize as $S_n = (n+\gamma)2\pi eB/\hbar$, where $\gamma$ is the phase offset which is 0 for Dirac fermions [34-36]. Close to the CNP, the constant-energy contour is isotropic, and $S_n$ can be simply expressed as $S_n = \pi k_n^2$. Thus, we have

$$k_n = \sqrt{2enB/\hbar}. \qquad (2)$$

The measured energy position $E_n$ and its associated $k_n$ for various $B$ then can trace out an effective band dispersion, as shown in Fig. 1(e). The corresponding $E_n$-$k_n$ exhibits a cubic dispersion relation as expected for ABC-stacked TLG. From the tight-binding calculation, we find $\gamma_0$ = 3.12 eV, $\gamma_1$ = 0.45 eV, and $\gamma_4$ = -0.102 eV is the best fit for the experimental data (see Fig. S2 in supporting materials [26] for more details). In our experiment, an electron-hole asymmetry [see Fig. 1(f) of the LL fan diagram] exceeds 17% is obtained (the effective mass of holes is larger than that of electrons). Such a large electron-hole asymmetry was also observed in bilayers [37,38] and in defective monolayer graphene [39]. In the ABC TLG, only the next-nearest interlayer hopping $\gamma_4$ among all the hopping parameters can account for the observed large electron-hole asymmetry. The deduced hopping parameters are consistent well with that obtained in theory [10] and in transport measurements [13].

With a close examination of the LL spectroscopy near the Fermi energy, we observe remarkable splittings of the degenerate LLs in $n = (0,1,2)$ and $n = 3$ levels. Each of the $n = 3$ LLs splits into two peaks in high magnetic fields [see Fig. 2(a) and Fig. S3 in supporting materials [26]]. Their splitting gaps for both the filled and the empty $n = 3$ LLs increase linearly with $B$ with similar slopes [Fig. 2(b)]. From the experimental data, we obtain the slope for the splitting as 0.35 meV/T or an effective $g$-factor of $g(LL_3) \approx 6$ with assuming a Zeeman-like dependence $E = g\mu_B B$ ($\mu_B$ is the Bohr magneton). The obtained effective $g$-factor is about three times of that for electron spins in graphene [40]. Therefore, we attribute the observed splitting of the $n = 3$ LLs to broken valley degeneracy. The valley degeneracy breaking is also observed in the $n = (0,1,2)$ LL. For the $n = (0,1,2)$ LL, the wave functions of electrons come from different valleys (+/-) are localized on the first and the third layer, respectively [31]. Such a layer polarization generates highly magnitude-asymmetric peaks of the valley splitting in the LL spectroscopy: $LL_{(0,1,2,+)}$ is much more intense than $LL_{(0,1,2,-)}$ in experiment because that the $LL_{(0,1,2,+)}$ mainly localized on the first layer and the STM tip predominantly probes the DOS of the top layer [21], as presented in Fig. 1(d) and Fig. 2(a). The valley splitting of the $n = (0,1,2)$ LL also exhibits a well linear scaling with $B$ [Fig. 2(b)]. However, we obtained a much larger slope, 1.34 meV/T or $g(LL_{(0,1,2)}) \approx 23$, than that of the $n = 3$ LLs. In a previous study, it was demonstrated that the valley splitting will increase (about 10%) when the measured LL in graphene monolayer is half filled [40]. In our experiment, the filling of the $n = (0,1,2)$ LL varies a lot in different magnetic fields [Fig. 2(a) and Fig. 2(c)], however, the valley splitting still follows an approximately linear scaling with the magnetic fields, indicating that the different filling of the $n = (0,1,2)$ LL cannot account for the large enhancement of the effective $g$-factor. The decrease of the effective $g$-factor with increasing LL index has also been observed in graphene monolayer [39]: the effective $g$-factor for the $n = 0$ LL is measured to be about 23 and it decreases to about 7 for the $n = 1$ LL. Such a behavior is attributed to exchange-driven quantum Hall ferromagnetism in graphene monolayer [41]. In the ABC TLG, the effective $g$-factor for the $n = (0,1,2)$ LL is nearly four times of that for the $n = 3$ LLs. Further theoretical works are needed to understand the approximately linear $B$-scaling of the valley splitting and the decrease of effective *g*-factor with increasing LL index in the ABC TLG.

The magnetic fields not only generate large valley splitting for the $n = (0,1,2)$ LL, but also further split the $LL_{(0,1,2,+)}$ and $LL_{(0,1,2,-)}$ LLs, as shown in Fig. 2(a). It is interesting to note that the intensity of one sub-peak is almost half of that of the other sub-peak in both the $LL_{(0,1,2,+)}$ and $LL_{(0,1,2,-)}$ (see Fig. S3 in supporting materials [26]). Because there is triplet orbital degeneracy for both the $LL_{(0,1,2,+)}$ and $LL_{(0,1,2,-)}$, therefore, it is reasonable to attribute the further splitting to broken orbital degeneracy. Figure 3 shows the measured gap size for orbital splittings of the $LL_{(0,1,2,+)}$ and $LL_{(0,1,2,-)}$ as a function of magnetic fields. In our experiment, the $LL_{(0,1,2,-)}$ is always unfilled. The orbital splitting of the $LL_{(0,1,2,-)}$ increases linearly with *B* and a slope of about 0.48 meV/T is obtained, as shown in Fig. 3(a). When the $LL_{(0,1,2,+)}$ is fully filled ($B > 10.5$ T), we also observe a linear increase of the orbital splitting with *B* and the slope is about 0.41 meV/T [Fig. 3(b)]. Previously, similar linear scaling orbital splitting with $\Delta \approx 0.1$ meV/T was observed in the lowest LL of the ABC TLG by transport measurements [16]. In the ABC-stacked TLG, the existence of a next-nearest interlayer hopping $\gamma_4$ is expected to lift the orbital degeneracy of the lowest LL and the orbital splitting in magnetic field can be expressed as, $\Delta = 3a^2 \gamma_0 \gamma_4 / \gamma_1 l_B^2$, where $a \approx 0.246$ nm is the lattice constant of graphene and $l_B \approx 26/\sqrt{B[T]}$ is the magnetic length [42]. Therefore, the orbital splitting should increase linearly with magnetic fields. By taking into account the obtained $\gamma_0$, $\gamma_1$, and $\gamma_4$ in our experiment (Fig. 1), we obtain $\Delta \approx 0.2$ meV/T, which is comparable to that measured in our experiment. The difference may mainly arise from the deviation of the value of $\gamma_4$ deduced from the electron-hole asymmetry of the LLs.

In our experiment, an obvious deviation from the linear scaling is also observed for 7 T< *B* < 10.5 T when the $LL_{(0,1,2,+)}$ is partially filled. In such a case, the orbital splitting of the $LL_{(0,1,2,+)}$ is much enhanced and reach a maximum when the Fermi level falls between the two split peaks at 9 T [Fig. 3(b)]. Such a result indicates that there are strong electron correlations in the studied system when the $LL_{(0,1,2,+)}$ is partially filled. Our experimental result suggests that electron correlations enhance the orbital splitting of the $LL_{(0,1,2,+)}$, but has negligible effect on the valley splitting of the $n = (0,1,2)$ LL (Figure 2). The interaction-enhanced orbital splitting in the partially filled $LL_{(0,1,2,+)}$ indicates that strongly correlated electron behaviour could emerge in the lowest LL of the ABC-stacked TLG. Recently, partial filling-induced Mott insulated states has been observed in the ABC TLG/hBN Moiré superlattice at the 1/4 and 1/2 fillings of the miniband [25], which correspond to one electron and two electrons per superlattice site, respectively. The highly degenerate flat bands of the ABC TLG in the quantum Hall regime can support more partial filling states to realize strongly many-body phenomena, such as charge density wave, fractional quantum Hall states, and quantum anomalous Hall states [19,42]. A single that might be related to such phenomena is that the energy splitting of the $LL_{(0,1,2,+)}$ shows a great (nearly double) enhancement when the Fermi level falls in the center of the $LL_{(0,1,2,+)}$, as shown in the spectrum at 9 T. Moreover, the two split sub-peaks at 9 T nearly have the same DOS integral, which quite differs from the expected result for the broken orbital degeneracy (see supporting materials [26] for more details). This phenomenon also has been observed in another ABC TLG sample (see Fig. S4 in supporting materials [26]). Although we still do not know the exact nature of this state, our result implies a new strongly correlated states emerges at half filling of the $LL_{(0,1,2,+)}$.

Further observations of strong electronic interactions in the ABC-stacked TLG are obtained by analysis of the quasiparticle lifetime. From the highly resolved LLs in our experiments, the quasiparticle lifetime $\tau$ can be extracted from the LL peak width $\Delta E_n$ with $\tau \approx \hbar/\Delta E_n$ (see Fig. S5 [26]). Figure 4 shows the energy dependence of the peak width (inverse lifetime) for filled $n > 3$ LLs (the split $LL_3$ and $LL_{(0,1,2,+)}$ are excluded to remove any possible effects of the valley and orbital splitting). Obviously, the peak width increases linearly with energy, yielding a charge-rate of the lifetime $\gamma' \sim 8$ fs/eV. In graphene system, the broadening of the LL peaks may originate from three possible scattering channels: (a) disorder, (b) electron-phonon coupling, and (c) electron-electron interactions. In our experiment, we can exclude the first two factors as the reason for the broadening of the LLs with increasing energy and such a phenomenon is attributed to electron-electron interactions (see supporting materials [26] for a detailed discussion). The interaction-induced linear energy scaling of the inverse lifetime has been observed in monolayer graphene [30] and graphite [44], which is treated as a clear feature of a non-Fermi liquid behavior [43]. The value of $\gamma'$ obtained here is similar as that ~ 9 fs/eV reported for graphene monolayer [30], indicating that the electronic interactions are important in the ABC TLG.

In summary, we have studied the electronic properties in the ABC-stacked TLG using LL spectroscopy. The linear magnetic-field-scaling of the valley splitting and orbital splitting are clearly observed. The valley splitting decreases dramatically with increasing the LL index. Moreover, we observe clear evidences for strong many-body effects, such as an obvious enhancement of the orbital splitting when the lowest LL is partially filled and linear energy scaling of the inverse lifetime of quasiparticles, in the ABC TLG. Our experiments indicate that there are many interesting broken-symmetry states and novel correlated phases in the quantum Hall regime of the ABC-stacked TLG.

**Acknowledgments**

This work was supported by the National Natural Science Foundation of China (Grant Nos. 11804089, 11674029, 11422430, 11374035), the Natural Science Foundation of Hunan Province, China (Grant No. 2018JJ3025), the National Basic Research Program of China (Grants Nos. 2014CB920903, 2013CBA01603). L.H. also acknowledges support from the National Program for Support of Top-notch Young Professionals, support from "the Fundamental Research Funds for the Central Universities", and support from "Chang Jiang Scholars Program". L.J.Y. acknowledges support from the Fundamental Research Funds for the Central Universities.

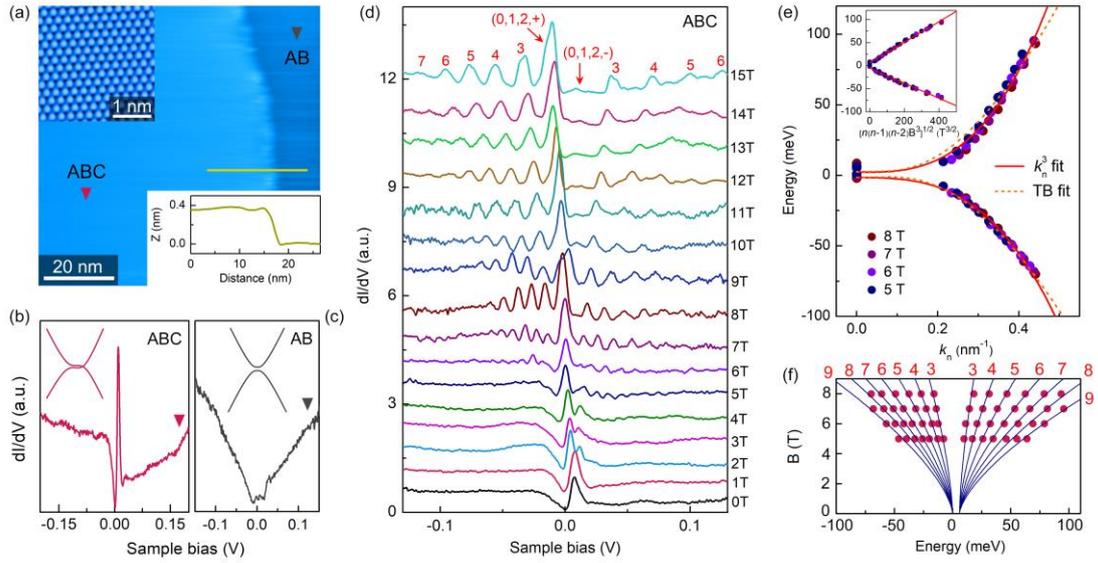

FIG. 1. Structural and spectroscopic characterization of the ABC-stacked TLG. (a) STM topographic image ($V_b$ = 100 mV, $I$ = 100 pA) of a decoupled ABC TLG on graphite. It shows a step edge in the right region of the image. Up-left inset: the atomic-resolution topographic image ($V_b$ = 50 mV, $I$ = 300 pA) of the ABC TLG surface showing the triangular contrasting. Bottom-right inset: height profile across the step along the solid line. The step height ~0.35 nm corresponding to one layer spacing. (b) and (c) Typical $dI/dV$ spectrum for ABC TLG and gaped AB bilayer recorded at the left and right region of the step in (a) at 0 T, respectively. The finite gap in AB region is generated by a substrate-induced interlayer potential. Insets show the corresponding band structures of the ABC TLG and gaped AB bilayer. (d) LL spectra of the ABC TLG recorded from 0 T to 15 T in 1 T increments. Curves are shifted vertically for clarity. (e) LL energies extracted from 5-8 T in (d) plotted versus momentum $k_n$ as deduced from Eq. (2). The solid and dashed lines are cubic and tight-binding (TB) fits to the data, respectively. The inset shows the $[n(n-1)(n-2)B^3]^{1/2}$ dependence of LL energies. (f) LL fan diagram under low fields. The red dots are data obtained from (d) and the blue curves are fits with Eq. (1). LLs up to $n$ = 9 are obtained for both electron and hole sides.

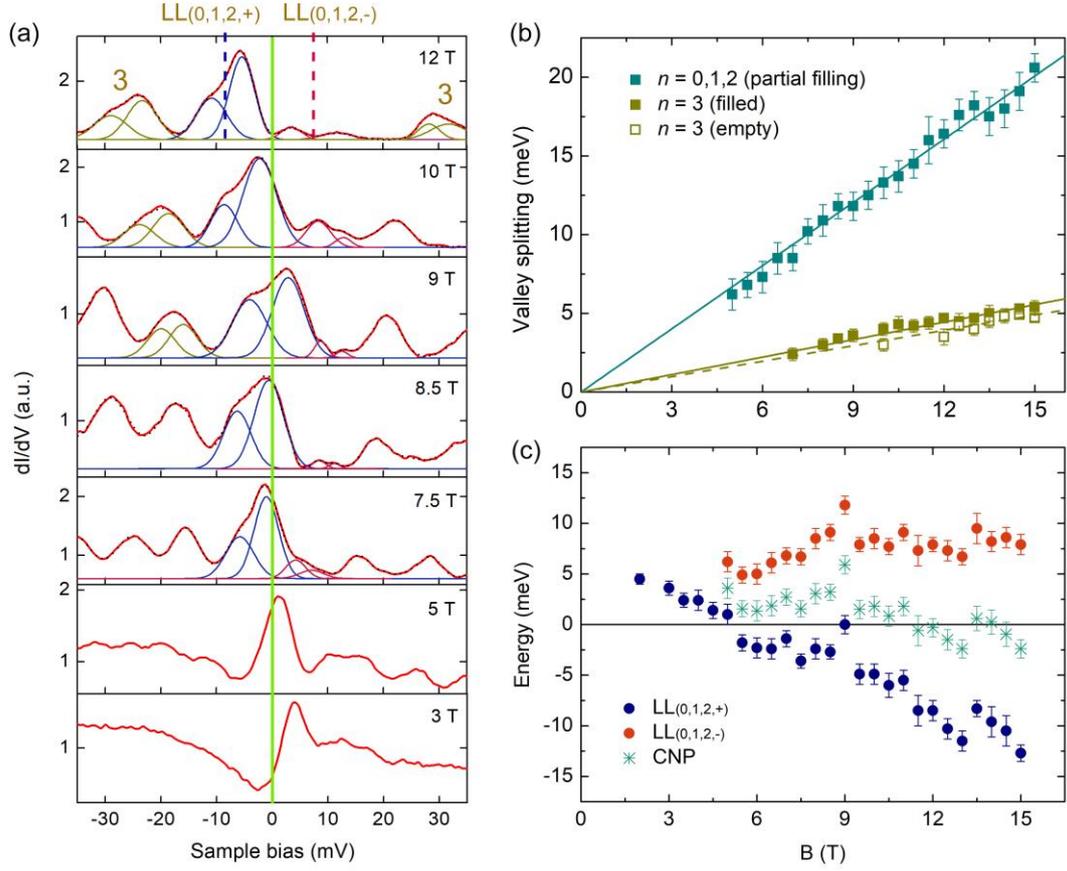

FIG. 2. Symmetry-broken level splittings and energy gaps of valley splitting. (a) $dI/dV$ spectra focusing on the Fermi level region for the indicated magnetic fields. The $n = (0,1,2)$ and $n = 3$ levels are all split in high fields (12 T as an example). $LL_3$ for both the filled and the empty states split into two peaks corresponding to the valley degeneracy broken. $LL_{(0,1,2)}$ splits into four peaks at the Fermi level with the larger splitting between $LL_{(0,1,2,+)}$ and $LL_{(0,1,2,-)}$ (marked by dashed lines in the top panel) corresponding to the lifting of valley degeneracy. The green line indicates the position of the Fermi level. (b) Valley splitting as a function of magnetic field for partially filled $LL_{(0,1,2)}$, filled and empty $LL_3$. Lines are linear fits to the data. (c) Magnetic field dependence of energy positions of $LL_{(0,1,2,+)}$, $LL_{(0,1,2,-)}$ and the CNP, showing different filling states of the $LL_{(0,1,2,+)}$ and $LL_{(0,1,2,-)}$.

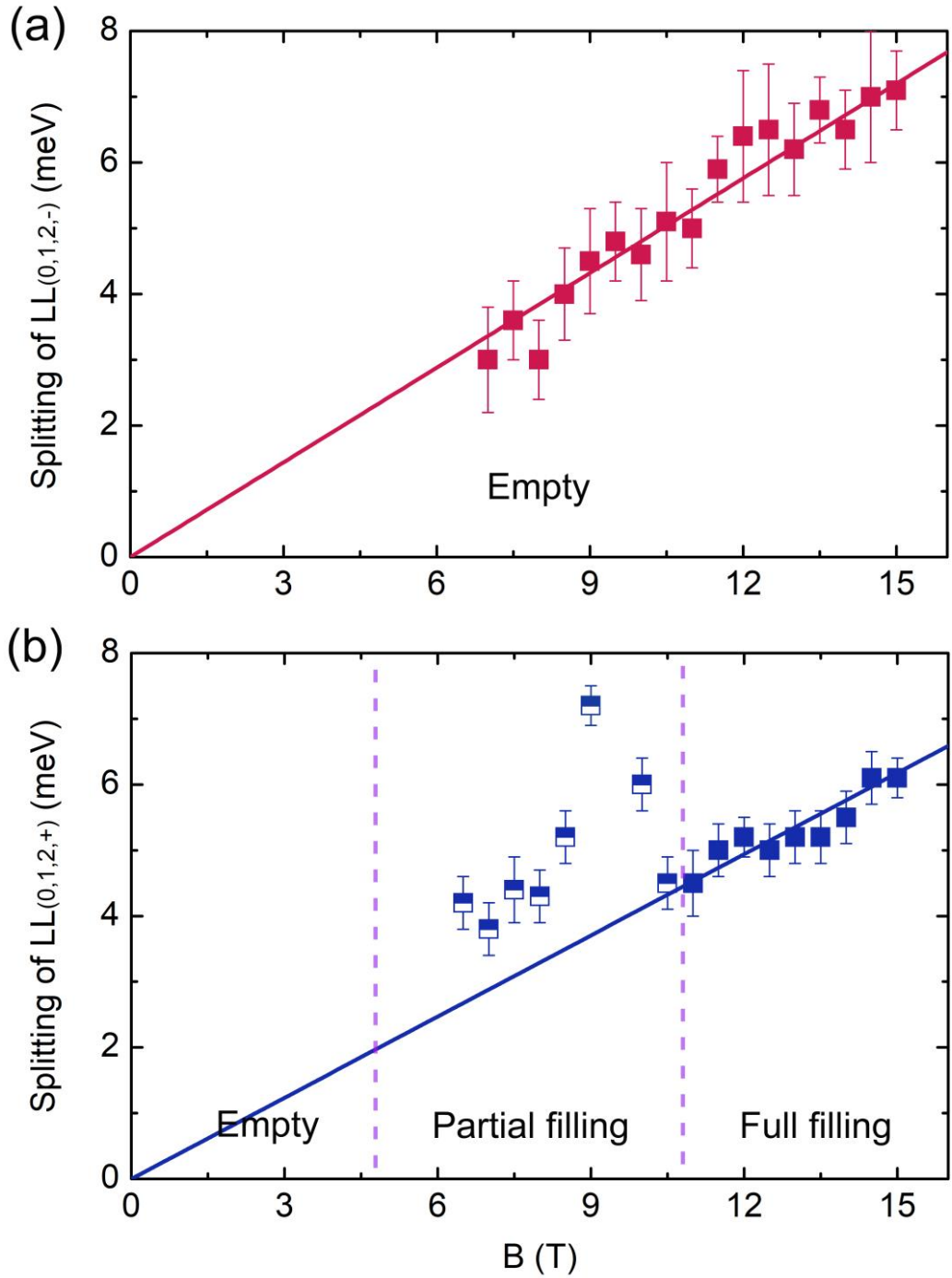

FIG. 3. Orbital splitting enhancement in partially filled $LL_{(0,1,2,+)}$. (a) Orbital splitting of the empty $LL_{(0,1,2,-)}$ as a function of magnetic field, revealing a linear scaling with the fit given by the solid line. (b) Splitting energy of $LL_{(0,1,2,+)}$ in partial and full fillings plotted versus magnetic field. The solid line is a reasonable linear fit to the data in full filling region.

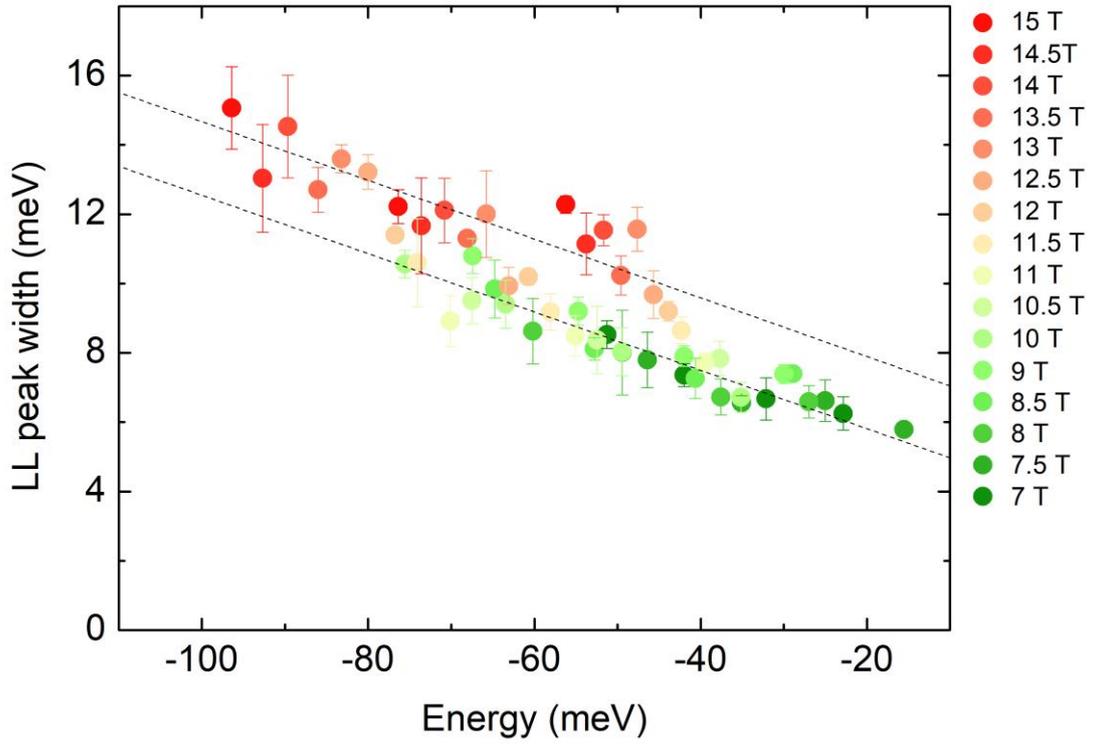

FIG. 4. Energy dependent lifetime of quasiparticles. Energy dependence of LL peak width, as deduced by Gaussian fit function, for *n* > 3 index at hole-side. The dashed lines are linear fits to the lower (≤ 10 T) and higher (≥ 12 T) fields with nearly the same slope ($\hbar/\gamma'$), showing the linear energy scaling and the field-induced increase of the peak width.